\documentclass[aip,jcp,reprint,twocolumn,superscriptaddress,showpacs,floatfix]{revtex4-2}
\usepackage{mathrsfs}
\usepackage{amssymb}
\usepackage{amsmath}
\usepackage[nodisplayskipstretch]{setspace}
\usepackage{parskip}

\usepackage{graphicx}  
\usepackage{braket}
\usepackage[dvipsnames]{xcolor}

\usepackage[colorlinks=true,citecolor=BrickRed]{hyperref}

\begin{document}
\title{A perturbative triples correction scheme to relativistic quadratic unitary coupled cluster method: Theory, implementation and benchmarking}
\author{Kamal Majee}
\affiliation{Department of Chemistry, Indian Institute of Technology Bombay, Powai, Mumbai 400076, India}
\author{J\'{a}n \v {S}imunek}
\affiliation{ Department of Inorganic Chemistry, Faculty of Natural Sciences, Comenius University, Bratislava
Ilkovičova 6, Mlynská dolina 842 15 Bratislava, Slovakia }
\author{Jozef  Noga}
\affiliation{ Department of Inorganic Chemistry, Faculty of Natural Sciences, Comenius University, Bratislava
Ilkovičova 6, Mlynská dolina 842 15 Bratislava, Slovakia }

\author{Achintya Kumar Dutta}
\thanks{Corresponding author}
\email[e-mail: ]{achintya.kumar.dutta@uniba.sk}
\affiliation{ Department of Inorganic Chemistry, Faculty of Natural Sciences, Comenius University, Bratislava
Ilkovičova 6, Mlynská dolina 842 15 Bratislava, Slovakia }

\begin{abstract}
We present a non-iterative triples correction to the relativistic quadratic unitary coupled cluster singles and doubles (qUCCSD) method, denoted as qUCCSD[T]. The method builds upon the Hermitian structure of the similarity transformed Hamiltonian in unitary coupled cluster method and can be derived using perturbational truncation to unitary coupled cluster energy functional. Relativistic effects are incorporated using the exact two-component atomic mean-field (X2CAMF) Hamiltonian, and the computational cost is further reduced through the frozen natural spinor (FNS) and Cholesky decomposition (CD) approximations. Benchmark results demonstrate that qUCCSD[T] outperforms previously proposed triples corrections to the unitary coupled cluster method and yields excellent agreement with experimental data and Full CI results. Additionally, the method shows high accuracy in computing the bond dissociation enthalpies, molecular geometries, vibrational frequencies, ionization potentials, and electron affinities of heavy-element-containing systems. The new qUCCSD[T] method is competitive to popular CCSD(T) even on a classical computer.  
\end{abstract}

\maketitle

\section{\label{sec:level1}Introduction\protect\\}

The accurate treatment of electron correlation remains one of the central challenges in quantum chemistry and chemical physics. The coupled cluster (CC) method,\cite{bartletbook_a, Cizek69_a} which uses an exponential parametrization of the wave function, has emerged as one of the most accurate and systematically improvable methods. The exponential parametrization guarantees the size extensivity of the energy, even in finite truncation of the cluster operator. The coupled cluster method is generally used in singles and doubles approximation of the cluster operator (CCSD). One can systematically improve the CCSD method by including operators of higher excitation ranks.\cite{Noga_fullCCSDT_a,Kucharski1986_a,Piecuchactivespace_a,MusialCCSDTQP_a} The CCSD(T) method\cite{full_ccsd_tNoga_a,CCSD_TRaghav_a} is particularly popular due to its inherent balance of computational cost and accuracy, and is considered the ``gold standard'' of quantum chemistry. \\ In addition to the standard coupled cluster method, the unitary coupled cluster (UCC)  variants\cite{NogaUCC_a,Kutzelnigg1977_a,PalUCC84_a,MDP85_a,Hoffmannucc_a} have attracted considerable attention\cite{Evangelistaucc_a} owing to their Hermitian nature and potential advantages for both classical and quantum computing platforms. The UCC method tends to converge\cite{KohnUCC_a} to  Full CI (FCI) results when higher excitation operators are considered.  However, unlike the standard CC method,  there is no natural truncation in Hamiltonian cluster commutator expansion in the UCC method, and it scales as FCI even with a finite truncation of the cluster operator.  Consequently, one cannot solve the UCC method on classical computers without evoking additional approximations.  One of the earliest adopted approximations relies on the perturbation theory-based truncation\cite{NogaUCC_a} of the UCC energy functionals. Perturbational approximation-based unitary coupled cluster method (UCC(n)), where n denotes the perturbation order) has been extended to triples and quadrupole excitation\cite{UCC5_a} operators, too. Bartlett and co-workers\cite{UCCSD_T_a} have recently reported a non-iterative triples correction scheme to the standard and perturbational approximation-based UCC method. However, the approximate UCC method based on the lower order of perturbation expansion often gives inferior results\cite{UCC_4_a,BartlettPT_a} compared to the standard CC method.\\
An alternative approach for the truncation of the UCC method is based on the commutator rank.\cite{Hoffmannucc_a,ChengUCC3_a} Among the various commutator-based truncation schemes of the UCC method, the quadratic unitary coupled cluster method(qUCC)\cite{qUCCSD1_a,quccsd2_a} has emerged as one giving the best compromise\cite{Phillips24062025} between computational cost and accuracy. The qUCC method is generally used in the singles and doubles approximation(qUCCSD)\cite{qUCCSD1_a} and has been implemented for ground and excited state energies.\cite{qUCCSD1_a,quccsd2_a,kamalucc_a} The natural orbital based low-cost implementation of qUCCSD and extension to the relativistic domain has also been achieved.\cite{kamalucc_a}  Yet, for systems where dynamic correlation is significant, particularly heavy-element systems, contributions from triple excitations are crucial for achieving chemical accuracy in addition to the relativistic effect\cite{vissherformulation_a}. Fully iterative inclusion of triple excitations within the qUCC is computationally prohibitive for all but the smallest systems, motivating the need for efficient perturbative treatments. At the same time, cost-effective treatment of the relativistic effect is necessary to make the method practical for real-life systems.
In this work, we develop a perturbative triple correction to the relativistic qUCCSD method, denoted as qUCCSD[T]. The exact two-component Hamiltonian with the atomic mean field treatment of the spin-orbit coupling effect (X2CAMF)\cite{x2camfcheng1_a,x2camfcheng2_a,x2camfripisky_a} has been used to introduce the relativistic effect.  Building upon the underlying structure of qUCCSD, we derive a consistent and computationally tractable perturbative framework to account for triple excitations. The paper is organized as follows: Section II presents the theoretical development of the perturbative triple correction within the relativistic qUCC framework. Section III provides benchmarking results on a set of molecular systems, followed by conclusions in Section IV.

\section{Theoretical framework}

\subsection{\label{app:subsec}Relativistic quadratic unitary coupled cluster theory}

In the UCC theory, the exact ground-state wave function $\ket{\Psi_{0}}$ is obtained by applying a unitary exponential operator $e^{\hat\sigma}$ to the reference wavefunction $\ket{\Phi_{0}}$, expressed as 
\begin{eqnarray}
\ket{\Psi_0}=e^{\hat{\sigma}}\ket{\Phi_0}
\label{eq:1}.
\end{eqnarray}

Where $\ket{\Phi_{0}}$ is generally, but not necessarily, a Dirac Hartree-Fock determinant.
The $\hat{\sigma}=\hat{T}-\hat{T}^{\dagger}$ denotes an anti-hermitian cluster operator, where $\hat{T}$ and $\hat{T}^{\dagger}$ denote the standard CC excitation and de-excitation operators, respectively.   The operator $\hat{\sigma}$ is generally restricted to one and two-body excitations, leading to the unitary coupled cluster singles and doubles (UCCSD) method.
\begin{eqnarray}
\hat{\sigma} = \hat{\sigma}_{1} + \hat{\sigma}_{2}
\label{eq:2}.
\end{eqnarray}
Where, 
\begin{eqnarray}
\hat{\sigma}_1 = \sum_{ia} \left[ {\sigma}_i^a \, \hat{c}_a^\dagger \hat{c}_i - ({\sigma}_i^a)^{*} \, \hat{c}_i^\dagger \hat{c}_a \right]
\label{eq:3}
\end{eqnarray}
\begin{eqnarray}
\hat{\sigma}_2 = \frac{1}{4} \sum_{ijab} \left[ {\sigma}_{ij}^{ab} \, \hat{c}_a^\dagger \hat{c}_b^\dagger \hat{c}_j \hat{c}_i - ({\sigma}_{ij}^{ab})^{*} \, \hat{c}_i^\dagger \hat{c}_j^\dagger \hat{c}_b \hat{c}_a \right]
\label{eq:4}
\end{eqnarray}
Inclusion of
\begin{align}
\hat{\sigma}_3 = \frac{1}{36} \sum_{ijkabc} \left[ {\sigma}_{ijk}^{abc} \, \hat{c}_a^\dagger \hat{c}_b^\dagger \hat{c}_c^\dagger \hat{c}_k\hat{c}_j \hat{c}_i 
- ({\sigma}_{ijk}^{abc})^{*} \, \hat{c}_i^\dagger \hat{c}_j^\dagger \hat{c}_k^\dagger\hat{c}_c\hat{c}_b \hat{c}_a \right]
\label{eq:5}
\end{align}
gives rise to the UCCSDT method.
where $i$, $j$, $k$, $l$ denote the occupied spinors, while $a$, $b$, $c$, $d$ represent virtual spinors. The quantities $\sigma_{i}^{a}$, $\sigma_{ij}^{ab}$ and $\sigma_{ijk}^{abc}$ are the singles, doubles and triples amplitudes, respectively. 
These cluster amplitudes($\sigma$) are determined by simultaneously solving a system of nonlinear equations
\begin{eqnarray}
\langle \Phi_{i}^{a} |\bar {H}| \Phi_0 \rangle =0
\label{eq:6}
\end{eqnarray}
\begin{eqnarray}
\langle \Phi_{ij}^{ab} |\bar {H}| \Phi_0 \rangle =0
\label{eq:7}
\end{eqnarray}
\begin{eqnarray}
\langle \Phi_{ijk}^{abc} |\bar {H}| \Phi_0 \rangle =0
\label{eq:8}
\end{eqnarray}
where
\begin{eqnarray}
\langle \Phi_{ijk...}^{abc...} | =\langle \Phi_0 | \hat{c}_i^\dagger \hat{c}_j^\dagger \hat{c}_k^\dagger \dots \hat{c}_c\hat{c}_b \hat{c}_a \dots
\label{eq:9}
\end{eqnarray}
and 
\begin{eqnarray}
\bar {H} =e^{-\hat{\sigma}}\hat{H}e^{\hat{\sigma}}
\label{eq:10}
\end{eqnarray}
A notable drawback of employing an anti-hermitian cluster operator in the UCC framework is that the Baker-Campbell-Hausdorff (BCH) expansion 
\begin{align}
\bar{H} = \hat{H} + [\hat{H}, \hat{\sigma}] 
+ \frac{1}{2!} [[\hat{H}, \hat{\sigma}], \hat{\sigma}] 
+ \frac{1}{3!} [[[\hat{H}, \hat{\sigma}], \hat{\sigma}], \hat{\sigma}] + \dots
\label{eq:11}
\end{align}
does not naturally truncate at finite order. Consequently, the practical application of UCC requires an artificial truncation of this infinite series.  However, a significant challenge lies in the lack of a universally preferred or unique scheme for this truncation. 
There are multiple strategies for truncating the BCH expansion in Eq. \ref{eq:11}. A popular approach involves applying arguments from many-body perturbation theory (MBPT).\cite{UCC_4_a,BartlettPT_a}
Alternatively, one can truncate the BCH expansion based on the depth of the commutator rank. One of the attractive schemes for commutator-based truncation of the similarity transformed Hamiltonian in the UCC method\cite{ChengUCC3_a} is based on the Bernoulli expansion.\cite{KutzlniggBernouli1_a, KutzlniggBernouli2_a} In this approach, the $\hat{H}$ is partitioned into the Fock operator ($\hat{F}$) and a fluctuation potential ($\hat{V}$) as
 \begin{eqnarray}
 \hat{H} = \hat{F} + \hat{V}.
 \label{eq:12}
 \end{eqnarray}
 The Fock operator is  block diagonal and rank-conserving operator for the  canonical Dirac Hartree-Fock method
  \begin{eqnarray}
 \hat{F} = \sum_{ij} f_{ij} \left\{ \hat{a}_i^{\dagger} \hat{a}_j \right\} + \sum_{ab} f_{ab} \left\{ \hat{a}_a^{\dagger} \hat{a}_b \right\}.
 \label{eq:13}
 \end{eqnarray}
 The fluctuation potential 
   \begin{eqnarray}
\hat{V} = \frac{1}{4} \langle pq || rs \rangle \left\{ \hat{a}_p^{\dagger} \hat{a}_q^{\dagger} \hat{a}_s \hat{a}_r \right\}.
 \label{eq:14}
 \end{eqnarray}
 can be further separated into the non-diagonal ($\hat{V}_{N}$) part which consist of pure excitation and de-excitation operator and the ``rest'' part ($\hat{V}_{R}$). Now, $\bar{H}$ can be expanded using Bernoulli numbers
\begin{eqnarray}
\bar{H}= \bar{H}^{0} + \bar{H}^{1} + \bar{H^{2}} +\bar{H^{3}} \dots
\label{eq:15}
\end{eqnarray}
with,
\begin{eqnarray}
\bar{H}^0 &=& \hat{F} + \hat{V} 
\label{eq:16}\\
%
\bar{H}^1 &=& [\hat{F}, \hat{\sigma}] + \frac{1}{2}[\hat{V}, \hat{\sigma}] + \frac{1}{2}[\hat{V}_R, \hat{\sigma}]
\label{eq:17}\\
\bar{H}^2 &=& \frac{1}{12}[[\hat{V}_N, \hat{\sigma}], \hat{\sigma}] + \frac{1}{4}[[\hat{V}, \hat{\sigma}]_R, \hat{\sigma}] \nonumber\\
&&+ \frac{1}{4}[[\hat{V}_R, \hat{\sigma}]_R, \hat{\sigma}]
\label{eq:18}\\
%
\bar{H}^{3}&=& \frac{1}{24}[[[\hat{V}_N, \hat{\sigma}], \hat{\sigma}]_R, \hat{\sigma}] + \frac{1}{8}[[[[\hat{V}_R, \hat{\sigma}]_R, \hat{\sigma}]_R, \hat{\sigma}] \nonumber\\
&& + \frac{1}{8}[[[\hat{V}, \hat{\sigma}]_R, \hat{\sigma}]_R, \hat{\sigma}] - \frac{1}{24}[[[\hat{V}, \hat{\sigma}]_R, \hat{\sigma}], \hat{\sigma}] \nonumber\\
&& - \frac{1}{24}[[[\hat{V}_R, \hat{\sigma}]_R, \hat{\sigma}], \hat{\sigma}].
\label{eq:19}
\end{eqnarray}

Such a structure not only simplifies the formulation but also provides a rigorous and efficient foundation for developing non-perturbative approximations within the UCC framework. One can derive an approximation to UCC method by taking the commutator up to a particular rank. For example, taking terms up to a $\bar{H}^3$ in the energy and $\bar{H}^2$ in the amplitude equation leads to a quadratic unitary coupled cluster approximation.\cite{qUCCSD1_a}
\begin{equation}   
E_{Gr}^{\text{qUCC}} = 
 \langle \Phi_0 | \bar{H}^{0}+\bar{H}^{1}+\bar{H}^{2}+\bar{H}^{3} | \Phi_0 \rangle 
\label{eq:20}
\end{equation}

\begin{align}
\langle \Phi_i^a | \bar{H}^0 + \bar{H}^1 + \bar{H}^2 | \Phi_0 \rangle =0 \nonumber\\
\langle \Phi_{ij}^{ab} | \bar{H}^0 + \bar{H}^1 + \bar{H}^2 | \Phi_0 \rangle = 0 \nonumber\\
 \langle \Phi_{ijk}^{abc} | \bar{H}^0 + \bar{H}^1 + \bar{H}^2 | \Phi_0 \rangle = 0 \nonumber\\
 \vdots \nonumber\\
\label{eq:21}
\end{align}
Taking $\hat{\sigma}=\hat{\sigma}_{1}+\hat{\sigma}_{2}$ and $\hat{\sigma}=\hat{\sigma}_{1}+\hat{\sigma}_{2}+\hat{\sigma}_{3}$ in equations \ref{eq:20} and \ref{eq:21} will lead to qUCCSD and qUCCSDT method, respectively.

\subsection{\label{app:ccsdt} Perturbative triples correction to qUCCSD method}
One of the most advantageous features of the qUCC method is that one can directly derive the perturbative triples correction from the perturbation-based truncation of the qUCCSDT energy expression in equation \ref{eq:20}. It is conceptually more straightforward than the CCSD(T) method in the standard coupled cluster method, where a rigorous derivation of the perturbative triples correction requires switching back and forth between projection and variational energy functional.\cite{CCSD_TRaghav_a,UCC5_a,StantonCCSD_T_a}
One can formulate the perturbative triples correction scheme to the qUCC method 
using MBPT, assuming a canonical Hartree-Fock (HF) reference state. Within this framework, the Fock operator $\hat{F}$ and the fluctuation operator $\hat{V}$ contribute at zeroth and first order in MBPT, respectively. The excitation operator $\hat{\sigma}_{n}$ typically emerges at the (n-1) order, with the exception of $\hat{\sigma}_{1}$, which contributes the second order. The qUCCSD method is complete up to fourth order in energy within singles and doubles truncation of the cluster operator.\cite{qUCCSD1_a} Additional terms arise due to the three-body operator, which will make the  energy expression complete up to the fourth order are 
\begin{eqnarray}
&&\Delta E_{qUCCSD[T]} = \nonumber \\
&&\quad\phantom{-} \frac{1}{8} \langle k j || a i \rangle \sigma_{a b c}^{l k j(2)} \sigma_{c b}^{i l}  
+\frac{1}{8} \langle i a || b c \rangle \sigma_{d a}^{j k} \sigma_{b c d}^{k j i(2)} \nonumber \\
 &&\quad -\frac{1}{8} \langle k a || i j \rangle \sigma_{c b a}^{j i l(2)} \sigma_{b c}^{l k} 
 -\frac{1}{8} \langle b a || c i \rangle \sigma_{d a b}^{i j k(2)} \sigma_{c d}^{k j}
\label{eq:22}
\end{eqnarray}
Where $\sigma_{a}^{i}$ and $\sigma_{ab}^{ij}$ are converged qUCCSD singles and doubles amplitudes. The $\sigma^{ijk(2)}_{abc}$ is the second-order triples amplitude

\begin{eqnarray}
    \sigma_{abc}^{ijk(2)}=\frac{s_{abc}^{ijk}}{\epsilon_{a}+\epsilon_{b}+\epsilon_{c}-\epsilon_{i}-\epsilon_{j}-\epsilon_{k}}
    \label{eq:23}
\end{eqnarray}
where
\begin{eqnarray}
s_{abc}^{i j k} &=&- \mathcal{P}(i j) \, \mathcal{P}(a b) \, \langle l a || j k \rangle \, \sigma_{b c}^{i l} 
- \mathcal{P}(a b) \, \langle l a || i j \rangle \, \sigma_{b c}^{k l} \nonumber\\
&&- \mathcal{P}(i j) \, \langle l c || j k \rangle \, \sigma_{a b}^{i l} 
- \langle l c || i j \rangle \, \sigma_{a b}^{k l} \nonumber\\
&&- \mathcal{P}(j k) \, \mathcal{P}(b c) \, \langle a b || d k \rangle \, \sigma_{d c}^{i j} - \mathcal{P}(b c) \, \langle a b || d i \rangle \, \sigma_{d c}^{j k} \nonumber\\
&&- \mathcal{P}(j k) \, \langle b c || d k \rangle \, \sigma_{d a}^{i j} 
- \langle b c || d i \rangle \, \sigma_{d a}^{j k}
\label{eq:24}
\end{eqnarray}
Here $\mathcal {P}$ is the permutation operator that swaps the indices
and $\epsilon$ are the diagonal elements of the Fock matrix.

\subsection{\label{app:subsec} Exact Two-component Hamiltonian with the atomic mean field  approximation }
One of the most rigorous ways to include the relativistic effect is to use a four-component Dirac Coulomb(4c-DC) Hamiltonian.\cite{dyall2007introduction_a} The high computational cost associated with 4c-DC Hamiltonian restricts its applicability beyond small to medium-sized systems. One of the practical ways to reduce the computational cost of relativistic calculations is to employ two-component Hamiltonians. Among the various flavors of two-component Hamiltonians available,\cite{hessRelativisticElectronicstructureCalculations1986_a,vanlentheRelativisticRegularTwocomponent1996_a,dyallInterfacingRelativisticNonrelativistic1997_a,nakajimaNewRelativisticTheory1999_a,baryszTwocomponentMethodsRelativistic2001_a,liuExactTwocomponentHamiltonians2009_a,saueRelativisticHamiltoniansChemistry2011_a,dyall2007introduction_a} we are going to use the X2CAMF approach.\cite{x2camfcheng1_a,x2camfcheng2_a,x2camfripisky_a} The 4c-DC Hamiltonian can be defined as
\begin{align}
\hat{H}^{4c}=\sum_{pq}h_{pq}^{4c}\hat{a}_{p}^{\dagger}\hat{a}_{q}^{\dagger}+\frac{1}{4}\sum_{pqrs}g_{pqrs}^{4c}\hat{a}_{p}^{\dagger}\hat{a}_{q}^{\dagger}\hat{a}_{s}\hat{a}_{r}
\label{eq:26}
\end{align}
Within the no-pair approximation,\cite{reiher2014relativistic,dyall2007introduction_a} the summation in the above equation is confined only to the positive-energy spinors. The indices $p$,$q$, $r$, and $s$ represented the positive-energy four-component spinors and the second quantized creation and annihilation operators are denoted by  $\hat{a}_{p}^{\dagger}$, $\hat{a}_{q}^{\dagger}$ and $\hat{a}_{r}$, $\hat{a}_{s}$ respectively. In the spin separation scheme, the matrix elements of two-electron interaction can be split into their spin-free (SF) and spin-dependent (SD) parts
\begin{eqnarray}
g_{pqrs}^{4c} = g_{pqrs}^{4c,SF} + g_{pqrs}^{4c,SD}
\label{eq:27}
\end{eqnarray}
Taking advantage of the localized nature of the spin-orbit interaction, the spin-dependent term can be approximated by the atomic mean field(AMF) approximation and the spin-free term can be approximated by the non-relativistic (NR) two-electron integrals 
\begin{align}
\label{eq:28}
\frac{1}{4}\sum_{pqrs}{g_{pqrs}^{\text{4c}}}a_{p}^{\dagger}a_{q}^{\dagger}a_{s}a_{r}\approx\sum_{pq}{g_{pq}^{\text{4c,AMF}}}a_{p}^{\dagger}a_{q}\nonumber \\+\frac{1}{4}\sum_{pqrs}{g_{pqrs}^{\text{NR}}}a_{p}^{\dagger}a_{q}^{\dagger}a_{s}a_{r}
\end{align}
By transforming into a two-component picture using the X2C transformation scheme, one arrives at an X2CAMF Hamiltonian, which contains an effective one-electron operator and the non-relativistic two-electron integrals. 
\begin{equation}
\label{eq:29}
\hat{H}^{\text{X2CAMF}}=\sum_{pq}{h_{pq}^{\text{X2CAMF}}}a_{p}^{\dagger}a_{q}+\frac{1}{4}\sum_{pqrs}{g_{pqrs}^{\text{NR}}}a_{p}^{\dagger}a_{q}^{\dagger}a_{s}a_{r}
\end{equation}
with 
\begin{equation}
\label{eq:30}
h^{\text{X2CAMF}}=h^{\text{X2C-1e}}+g^{\text{2c,AMF}}
\end{equation}
One of the most prominent advantages of this approach is that the required two electrons are only non-relativistic. Therefore, one can use the already well-established techniques for the efficient treatment of non-relativistic two-electron integrals in the relativistic calculations. In the present work, the two-electron integrals are treated using the Cholesky decomposition(CD) technique.\cite{helmich-parisRelativisticCholeskydecomposedDensity2019_a,banerjeeRelativisticResolutionoftheidentityCholesky2023_a,uhlirovaCholeskyDecompositionSpinFree2024_a,zhangCholeskyDecompositionBasedImplementation2024_a} 
\subsection{\label{app:ccsdt} Cholesky Decomposition}
In the CD framework,\cite{beebe1977simplifications} the two-electron repulsion integrals (ERI) can be efficiently approximated as,
\begin{align}
(\mu \nu \mid \kappa \lambda) \approx \sum_{P}^{n_{CD}}L_{\mu \nu}^{P}L_{\kappa \lambda}^{P}
\label{eq:31}
\end{align}
where $\mu$, $\nu$, $\kappa$, $\lambda$ correspond to atomic spinor indices, $L_{\mu \nu}^{P}$ denotes the Cholesky vectors (CVs) and $n_{CD}$ specifies their dimensionality. Both single-step and double-step algorithms\cite{aquilante2011cholesky,folkestad2019efficient,zhang2021toward} have been proposed to perform CD of ERIs efficiently. In this present work, we employ the standard single-step approach, in which CVs are generated iteratively by selecting the largest diagonal elements of the ERI matrix $(\mu \nu \mid \mu \nu)$. The procedure is repeated until the largest diagonal elements fall below a predefined Cholesky threshold $\tau$, which determines the accuracy of the resulting decomposition.

The computed CVs on the AO basis can be transformed to the MO basis as follows,

\begin{align}
L_{pq}^{P}=\sum_{\mu\nu}C_{\mu p}^{*}L_{\mu \nu}^{P}C_{\nu q}
\label{eq:32}
\end{align}
Using these transformed vectors, anti-symmetrized two electron integrals can be efficiently generated on the fly within the molecular orbital (MO) basis.
\begin{align}
\langle p q|| r s \rangle = \sum_{p}^{n_{CD}}(L_{pr}^{P}L_{qs}^{P} - L_{ps}^{P}L_{qr}^{P}) 
\label{eq:33}
\end{align}  In the present implementation, the integrals involving two or fewer virtual indices are explicitly constructed and stored. Whereas, the integrals of the forms $\langle a b|| c d \rangle$ and $\langle a b|| c i \rangle$ are computed on the fly.

\subsection{\label{app:subsec}Frozen Natural Spinors (FNS)}
Relativistic CC calculations are more expensive than their non-relativistic counterparts due to the loss of spin symmetry and the additional need to store and process complex-valued quantities. Moreover, one often needs to uncontract the basis set to fit the small component of the Hamiltonian in relativistic calculations, which further increases the cost of relativistic CC calculations. One of the most effective ways to reduce the computational cost of relativistic CC calculations is to truncate the virtual space using natural spinors.\cite{chamoliReducedCostFourcomponent2022_a, yuanAssessingMP2Frozen2022_a} The natural spinors are the relativistic analogs of the natural orbitals introduced by Löwdin\cite{lowdinQuantumTheoryManyParticle1955_a} and can be obtained by diagonalizing the one-body correlated reduced density matrix (1-RDM), which is constructed from a spin-orbit coupled wave function calculated through some approximate electron correlation calculations.\cite{chamoliRelativisticReducedDensity2024}  Among the various flavors of natural spinors,\cite{SSFNS_a, PSFNS_a,chamoliReducedCostFourcomponent2022_a} the MP2-based natural spinors are the preferred choice for ground-state coupled cluster correlated energy calculations due to their favorable computational cost. \\
The construction of natural spinors using the MP2 method can be performed as follows. The virtual-virtual block of 1-RDM $(\mathbf{D}^\mathrm{(vv)})$ in the MP2 method can be expressed as 
\begin{equation}
D_{ab}^{\mathrm{(vv)}} = -\frac{1}{2} \sum_{c,i > j}\frac{\langle ac||ij\rangle {\langle bc||ij\rangle}}{\epsilon_{ij}^{ac}\epsilon_{ij}^{bc}}
\label{eq:34}
\end{equation}
Where, 
\begin{equation}
\epsilon_{ij}^{dc}=\epsilon_i+\epsilon_j-\epsilon_d-\epsilon_c\,, \quad d=a,b
\label{eq:35}
\end{equation}
in Eqs. \ref{eq:35}, the molecular spinor energies are defined by $\epsilon_i$, $\epsilon_j$, $\epsilon_a$, $\epsilon_b$, and $\epsilon_c$, whereas $\langle dc||ij\rangle$ $(d$=$a,b)$ 
denotes the antisymmetrized two-electron integrals. 
One can diagonalize the $\mathbf{D}^{\mathrm{(vv)}}$
\begin{equation}
\mathbf{D}^{\mathrm{(vv)}}\mathcal{V}=n{\mathcal{V}}
\label{eq:36}
\end{equation}
The eigenvectors $(\mathcal{V})$ are referred to as virtual natural spinors, while the corresponding eigenvalues $(n)$ are known as occupation numbers. These spinors are arranged according to their contributions to the total correlation energy, as indicated by their occupation numbers. By introducing a predefined threshold for the occupation number, the virtual space can be systematically truncated, retaining only those spinors whose occupancy exceeds the cut-off. After this selection, the virtual-virtual block of the Fock matrix is transformed into the truncated basis of the selected natural spinors
\begin{equation}
F_{\mathcal{VV}}^{NS}={\tilde{\mathcal V}}^\dagger F_{vv}\tilde{\mathcal V}
\label{eq:37}
\end{equation}
Where $\tilde{\mathcal V}$ refers to the virtual natural spinors defined within a truncated basis set, while $F_{vv}$ corresponds to the virtual-virtual block of the initial canonical Fock matrix. To obtain the semi-canonical virtual natural spinors,  $F_{\mathcal{VV}}^{NS}$ is diagonalized  

\begin{equation}
F_\mathcal{VV}^{NS}\tilde{\mathcal Z}=\epsilon\tilde{\mathcal Z}
\label{eq:38}
\end{equation}
The transformation matrix $(\mathcal B)$ transforms the canonical virtual spinor space to the semi-canonical natural virtual spinor space
\begin{equation}
\mathcal B=\tilde{\mathcal V}\tilde{\mathcal Z}
\label{eq:39}
\end{equation}

Consequently, the employed basis set contains the canonical occupied spinors along with semi-canonical virtual natural spinors and is denoted as the frozen natural spinors (FNS) approximation. The use of FNS can significantly reduce the computational cost of relativistic coupled cluster calculations.  One can correct for the truncated virtual space in the qUCC method perturbatively as 
\begin{eqnarray}
 E_{qUCC}^{canonical}={E_{qUCC}^{FNS}}+\Delta E_{qUCC} 
 \label{eq:40}
\end{eqnarray}
\begin{eqnarray}
\Delta E_{qUCC} \approx \Delta E_{MP2}
\label{eq:41}
\end{eqnarray}
so, 
\begin{eqnarray}
\Delta E_{MP2} = E_{MP2}^{canonical} - E_{MP2}^{FNS}
\label{eq:42}
\end{eqnarray}

\subsection{Implementation and Computational details}
The new relativistic qUCCSD[T] method has been implemented into the development version of our in-house quantum chemistry software package BAGH.\cite{duttaBAGHQuantumChemistry2025_a} The package is written primarily in Python while the performance-critical components are optimized using Cython and Fortran. BAGH is interfaced with PYSCF,\cite{sunLibcintEfficientGeneral2015_a, sunPySCFPythonbasedSimulations2018_a, sunRecentDevelopmentsSCF2020_a} GAMESS US,\cite{barcaRecentDevelopmentsGeneral2020_a} socutils,\cite{wangXubwaSocutils2025_a} and Dirac.\cite{saue_2025_14833106} The X2CAMF-HF calculations are carried out using the socutils package,\cite{wangXubwaSocutils2025_a} interfaced with BAGH. More details about the CD-X2CAMF-based implementation of the relativistic coupled cluster method can be found in Ref.~\onlinecite{chamoli2025frozen_a}. Three predefined settings of the FNS and CD thresholds can be used: LOOSEFNS (FNS threshold: $10^{-4}$, and CD threshold: $10^{-3}$), NORMALFNS (FNS threshold: $10^{-4.5}$, and CD threshold: $10^{-4}$), and TIGHTFNS (FNS threshold: $10^{-5}$, and CD threshold: $10^{-5}$).The LOOSEFNS setting has been found to be appropriate for rapid estimation, normal FNS for standard energy difference, and the TIGHTFNS setting has been found to be appropriate for finite-field calculation of properties.\cite{chamoli2025frozen_a} 
\section{Result and Discussion}
\subsection{comparison with full CI and other triples correction schemes to UCC method in the non-relativistic regime} 
\begin{figure*} 
\includegraphics[width=1.0\textwidth]{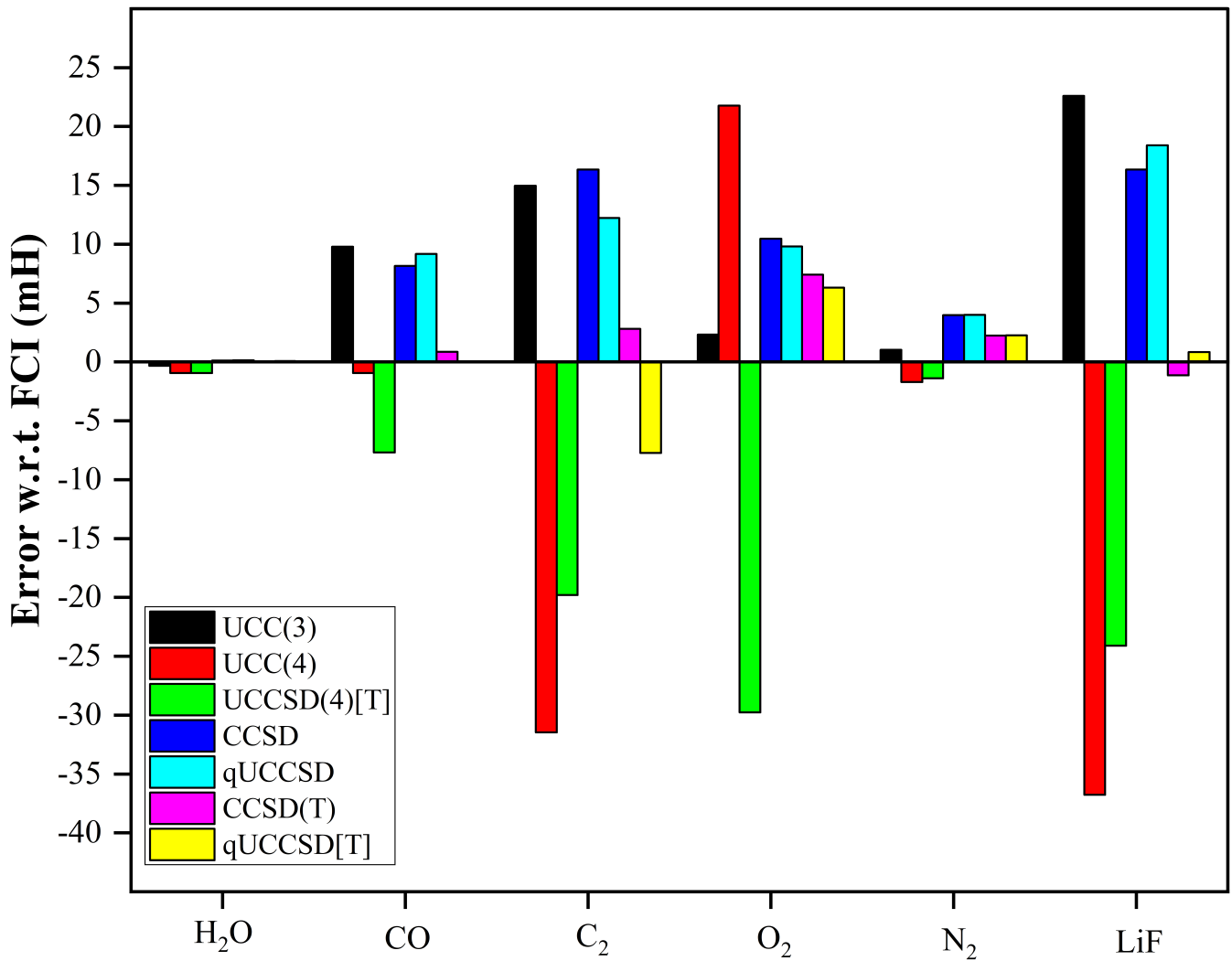} 
\caption{Deviation in total molecular energies (in mH) from FCI results using the frozen core approximation for various approximate unitary and standard coupled cluster methods.}
\label{fig:my_FCI}
\end{figure*}

Bartlett and co-workers\cite{UCCSD_T_a} have recently compared the various triples correction schemes to UCC methods with FCI results both for classical and quantum computers. As the present implementation of qUCCSD$[T]$ is restricted to classical computers only, we have chosen to compare with the variants that can be efficiently solved on classical computers. 
\hbox{STO-6G} basis set was used for the calculations together with the frozen core approximation following Ref.~\onlinecite{UCCSD_T_a}. The UCC(3), UCC(4), UCCSD(4)$[T]$, CCSD, CCSD(T) and FCI results are taken from 
Ref.~\onlinecite{UCCSD_T_a}. Figure \ref{fig:my_FCI} shows a schematic description of the error with respect to the FCI results, and the corresponding values are presented in table \ref{table:FCI}. The errors are reported in milli Hartree(mH). CD and FNS approximations have not been used for the non-relativistic calculations.
Although it can be seen that UCC(3), qUCCSD, and CCSD give very similar performance as compared to the FCI method, the performance of the partial triples correction scheme shows considerable deviation among each other. The UCC(4) shows the worst performance with an error as high as -36.78 mH. The UCCSD(4)$[T]$  method gives slightly better performance with a maximum error of 29.77 mH. The newly developed  qUCCSD$[T]$ method gives the best performance with a maximum error of - 7.728 mH, and its performance is comparable to the standard CCSD(T) method.

\begin{table*}[h]
\caption {Errors (mH) with respect to FCI  calculated using various methods in \textsc{STO}-6G basis set with the frozen core approximation.}
\begin{ruledtabular}
\begin{tabular}{ccccccccc}
Molecule & FCI\cite{UCCSD_T_a} & UCC(3)\cite{UCCSD_T_a} & UCC(4)\cite{UCCSD_T_a} & UCCSD(4)[T]\cite{UCCSD_T_a} & CCSD\cite{UCCSD_T_a} & qUCCSD & CCSD(T)\cite{UCCSD_T_a} & qUCCSD[T]  \\ \hline

$H_{2}O$ & -75.7287768   & -0.338  & -0.942  & -0.936  & 0.118  & 0.139   & 0.05   &  0.062 \\
CO       & -112.4426091  &  9.795  & -0.955  & -7.697  & 8.157  & 9.162   & 0.865  & -0.013 \\
$C_{2}$  & -75.4339744   &  14.97  & -31.46  & -19.81  & 16.33  & 12.228  & 2.817  & -7.728 \\
$O_{2}$  & -149.1251956  &  2.312  &  21.77  &  29.77  & 10.47  & 9.816   & 7.411  &  6.315 \\
$N_{2}$  & -108.7003854  &  1.036  & -1.707  & -1.396  & 3.983  & 3.986   & 2.231  &  2.251 \\
LiF      & -106.4434636  &  22.59  & -36.78  & -24.11  & 16.33  & 18.396  &-1.142  &  0.838\\
    
\end{tabular}
\end{ruledtabular}
\label{table:FCI}
\end{table*}

\subsection{Dissociation Enthalpy of Heavy metal-ligand complexes}
\begin{figure*} 
\includegraphics[width=1.0\textwidth]{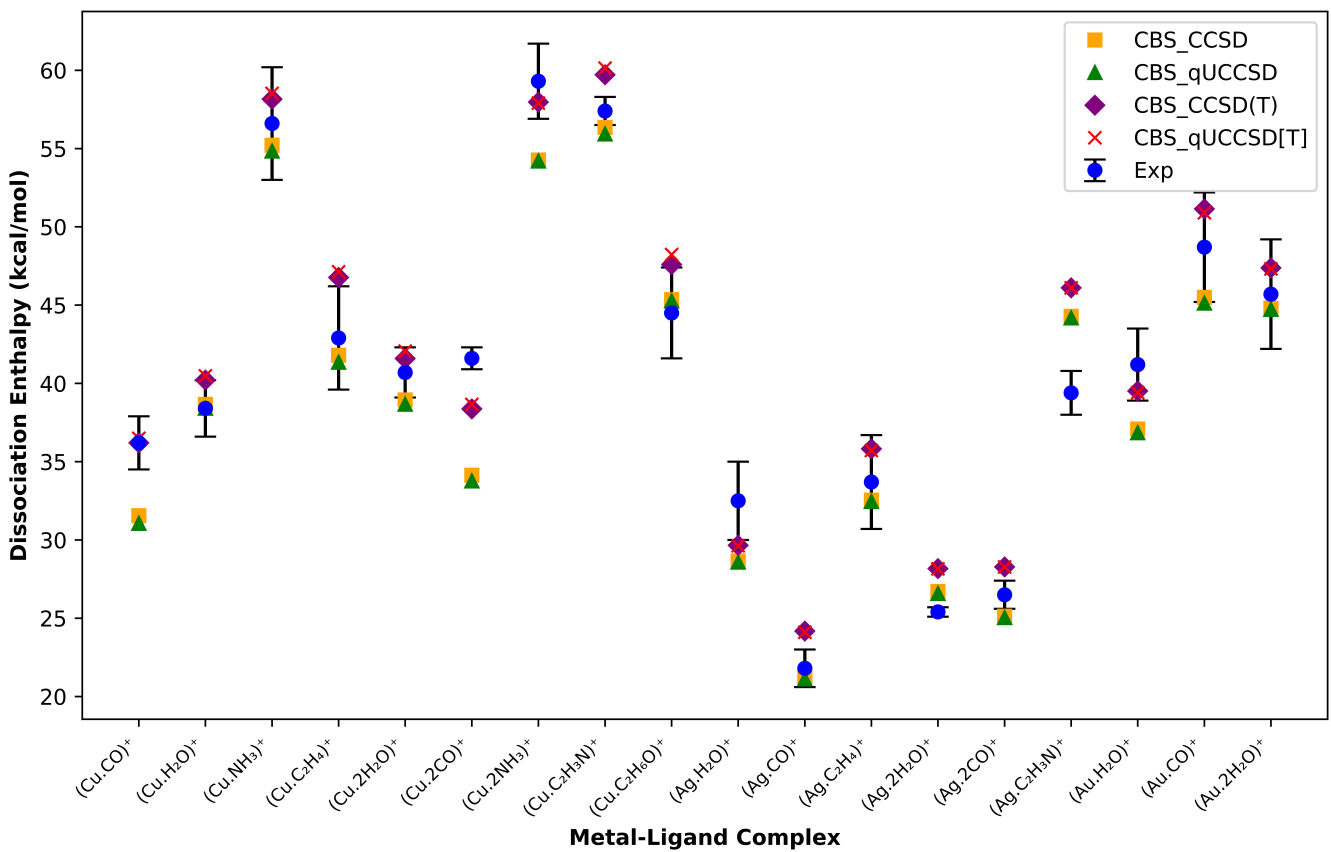} 
\caption{Experimental bond dissociation enthalpies (with error bars), along with those computed using the FNS-CD-X2CAMF-based CCSD, qUCCSD, CCSD(T), and qUCCSD[T] method at complete basis set (CBS) limit.}
\label{fig:my_BDE}
\end{figure*}

To evaluate the practical applicability and accuracy of the qUCCSD[T] method in the relativistic domain, we computed the bond dissociation enthalpies (BDEs) of a representative set of 18 coinage metal-ligand complexes involving Cu$^+$, Ag$^+$, and Au$^+$ ions. This test set was taken from Ref.~\onlinecite{chamoli2025frozen_a}. These systems are particularly challenging due to the combined influence of strong electron correlation and relativistic effects. All calculations were performed  using the FNS-CD-X2CAMF framework, as detailed in Section IIF, and the NORMALFNS truncation threshold has been used for the calculation. The optimized geometries for all systems were taken from the study by Cavallo and co-workers.\cite{Cavallo_geometry} For the ligand atoms, the aug-cc-pVXZ(X=D,T and Q) has been used. On the other hand, the dyall.aexz basis sets (x = 2, 3, and 4) were used for the metal cations Cu$^{+}$, Ag$^{+}$, and Au$^{+}$. All basis sets were used in their uncontracted forms, and the frozen core approximation was used. The calculated HF and correlation energies were extrapolated using the three-point extrapolation by Peterson and Dunning.\cite{peterson} \\
The computed BDEs from qUCCSD[T] are compared against those from CCSD, CCSD(T), and qUCCSD in Table \ref{table:BDE}, while Fig. \ref{fig:my_BDE} provides a visual comparison with experimental values and associated error bars. It is evident that the inclusion of perturbative triples through qUCCSD[T] leads to a systematic improvement over the qUCCSD method, often providing results that are closely aligned with the experimental reference data. For example, in the case of $(Cu\cdot CO)^{+}$, the qUCCSD[T] value of 36.49 kcal/mol is in excellent agreement with the experimental value of 36.2 ± 1.7 kcal/mol. A similar level of agreement is observed across other systems, such as $(Ag\cdot CO)^{+}$ and $(Au\cdot H_{2}O)$, further validating the robustness of the method.
Importantly, qUCCSD[T] consistently outperforms qUCCSD by accounting for the missing dynamic correlation from triple excitations, which is crucial for metal–ligand interactions involving heavier elements. The method also achieves accuracy on par with the standard CCSD(T).

\begin{table*}[h]
\caption{Experimental bond dissociation enthalpies (kcal/mol) for metal-ligand complexes  (with error bars), along with those computed using the FNS-CD-X2CAMF based CCSD, qUCCSD, CCSD(T) and qUCCSD[T]] at complete basis set (CBS) limit.}
\begin{ruledtabular}
\begin{tabular}{cccccc}
Reactions & CBS-CCSD\cite{chamoli2025frozen_a} & CBS-qUCCSD & CBS-CCSD(T)\cite{chamoli2025frozen_a} & CBS-qUCCSD[T] & Exp.\cite{meyer1995sequential,dalleska1994solvation,walter1998sequential,sievers1998transition,koizumi2001collision,el2002binding,guo1991bonding,holland1982thermochemical,shoeib2001silver,schwarz2003relativistic,poisson2002multifragmentation,vitale2001solvation} \\ \hline
\(\mathrm{(Cu\cdot CO)}^+ \rightarrow \mathrm{Cu}^+ + \mathrm{CO}\)                    & 31.55 & 31.06 & 36.20 & 36.49 & 36.2 ± 1.7 \\
\(\mathrm{(Cu\cdot H_{2}O)}^+ \rightarrow \mathrm{Cu}^+ + \mathrm{H_{2}O}\)            & 38.66 & 38.41 & 40.22 & 40.48 & 38.4 ± 1.8 \\
\(\mathrm{(Cu\cdot NH_{3})}^+ \rightarrow \mathrm{Cu}^+ + \mathrm{NH_{3}}\)            & 55.21 & 54.84 & 58.16 & 58.52 & 56.6 ± 3.6 \\
\(\mathrm{(Cu\cdot C_{2}H_{4})}^+ \rightarrow \mathrm{Cu}^+ + \mathrm{C_{2}H_{4}}\)    & 41.80 & 41.35 & 46.77 & 47.13 & 42.9 ± 3.3 \\
\(\mathrm{(Cu\cdot 2H_2O)}^+ \rightarrow \mathrm{(Cu\cdot H_2O)}^+ + \mathrm{H_2O}\)   & 38.94 & 38.67 & 41.59 & 42.06 & 40.7 ± 1.6 \\
\(\mathrm{(Cu\cdot 2CO)}^+ \rightarrow \mathrm{(Cu\cdot CO)}^+ + \mathrm{CO}\)         & 34.13 & 33.77 & 38.37 & 38.66 & 41.6 ± 0.7 \\
\(\mathrm{(Cu\cdot 2NH_3)}^+ \rightarrow \mathrm{(Cu\cdot NH_3)}^+ + \mathrm{NH_3}\)   & 54.27 & 54.20 & 57.97 & 57.90 & 59.3 ± 2.4 \\
\(\mathrm{(Cu\cdot C_{2}H_{3}N)}^+ \rightarrow \mathrm{Cu}^+ + \mathrm{C_{2}H_{3}N}\)  & 56.35 & 55.93 & 59.72 & 60.14 & 57.4 ± 0.9 \\
\(\mathrm{(Cu\cdot C_{2}H_{6}O)}^+ \rightarrow \mathrm{Cu}^+ + \mathrm{C_{2}H_{6}O}\)  & 45.35 & 45.27 & 47.59 & 48.23 & 44.5 ± 2.9 \\
\(\mathrm{(Ag\cdot H_{2}O)}^+ \rightarrow \mathrm{Ag}^+ + \mathrm{H_{2}O}\)            & 28.66 & 28.58 & 29.66 & 29.63 & 32.5 ± 2.5 \\
\(\mathrm{(Ag\cdot CO)}^+ \rightarrow \mathrm{Ag}^+ + \mathrm{CO}\)                    & 21.28 & 21.12 & 24.18 & 24.10 & 21.8 ± 1.2 \\
\(\mathrm{(Ag\cdot C_{2}H_{4})}^+ \rightarrow \mathrm{Ag}^+ + \mathrm{C_{2}H_{4}}\)    & 32.54 & 32.46 & 35.82 & 35.71 & 33.7 ± 3.0 \\
\(\mathrm{(Ag\cdot 2H_2O)}^+ \rightarrow \mathrm{(Ag\cdot H_2O)}^+ + \mathrm{H_2O}\)   & 26.71 & 26.58 & 28.17 & 28.15 & 25.4 ± 0.3 \\
\(\mathrm{(Ag\cdot 2CO)}^+ \rightarrow \mathrm{(Ag\cdot CO)}^+ + \mathrm{CO}\)         & 25.13 & 25.04 & 28.28 & 28.28 & 26.5 ± 0.9 \\
\(\mathrm{(Ag\cdot C_{2}H_{3}N)}^+ \rightarrow \mathrm{Ag}^+ + \mathrm{C_{2}H_{3}N}\)  & 44.28 & 44.18 & 46.10 & 46.10 & 39.4 ± 1.4 \\
\(\mathrm{(Au\cdot H_{2}O)}^+ \rightarrow \mathrm{Au}^+ + \mathrm{H_{2}O}\)            & 37.09 & 36.85 &39.51  & 39.41 & 41.2 ± 2.3 \\
\(\mathrm{(Au\cdot CO)}^+ \rightarrow \mathrm{Au}^+ + \mathrm{CO}\)                    & 45.49 & 45.13 & 51.15 & 50.90 & 48.7 ± 3.5 \\
\(\mathrm{(Au\cdot 2H_2O)}^+ \rightarrow \mathrm{(Au\cdot H_2O)}^+ + \mathrm{H_2O}\)   & 44.80 & 44.73 & 47.38 & 47.31 & 45.7 ± 3.5 \\

\end{tabular}
\end{ruledtabular}
\label{table:BDE}
\end{table*}

\subsection{Bond Length and Vibrational Frequency} 
To further assess the performance of the relativistic qUCCSD[T] method, we benchmarked its ability to predict equilibrium bond lengths and harmonic vibrational frequencies for a series of diatomic molecules (HF, HCl, HBr, and HI), incorporating both light and heavy elements. The calculations were performed using uncontracted aug-cc-pVTZ (tz) and aug-cc-pVQZ (qz) basis sets for lighter atoms (H, F, Cl) and dyall.acv3z/dyall.acv4z for heavier atoms (Br, I).The bond length and harmonic vibrational frequency are calculated by numerical differentiation of the total energy using the TWOFIT utility program of DIRAC.\cite{saue_2025_14833106} A fifth-order polynomial is used, and the TIGHTFNS threshold has been used for the calculations.

Table \ref{table:BL} presents the bond lengths obtained from CCSD, CCSD(T), qUCCSD, and qUCCSD[T], compared to experimental values.\cite{huber2013molecular_a} The change from tz to qz level is small, so one could consider the qz results to be almost converged with respect to the basis set. The qUCCSD[T] method yields bond distances that are in excellent agreement with experiment and nearly indistinguishable from those predicted by CCSD(T), in addition to the advantage of its Hermitian formulation and consistent inclusion of spin–orbit effects via the X2CAMF Hamiltonian.

In Table \ref{table:hvib}, we report the corresponding harmonic vibrational frequencies. The trends observed in bond lengths are mirrored here: qUCCSD[T] reproduces experimental vibrational frequencies with high fidelity, typically within a few wavenumbers. For example, the qUCCSD[T] frequency for HF is 4138.25 cm$^{-1}$, compared to the experimental value of 4138.32 cm$^{-1}$, showing high accuracy. Across the dataset, qUCCSD[T] performs on par with CCSD(T), slightly outperforming both CCSD and qUCCSD.

These results indicate that the qUCCSD[T] method is not only reliable for total energy and enthalpy calculations but also provides accurate molecular geometries and vibrational properties, making it a comprehensive approach for studying both light and heavy-element-containing molecules.

\begin{table*}[ht!]
\caption{Comparison of bond length ($\AA$) calculated using the UCC-based method (qUCCSD and qUCCSD[T]) and the normal coupled cluster method(CCSD and CCSD(T)), under a fixed FNS and CD threshold $10^ {-5}$ with frozen core approximation.}.
\begin{ruledtabular}
\begin{tabular}{cccccccccc}
Molecule & \multicolumn{2}{c} {CCSD} & \multicolumn{2}{c}{qUCCSD} & \multicolumn{2}{c}{CCSD(T)} &\multicolumn{2}{c}{qUCCSD[T]} & Exp.\cite{huber2013molecular_a} \\
\cline{2-3}
\cline{4-5}
\cline{6-7}
\cline{8-9}
& tz & qz & tz & qz& tz& qz & tz & qz \\ \hline
HF  & 0.9178 & 0.9151 & 0.9177 & 0.9149 & 0.9207 & 0.9179 & 0.9205 & 0.9178 & 0.9168  \\

HCl & 1.2760 & 1.2751  & 1.2760 & 1.2751 & 1.2782 & 1.2774 & 1.2781 & 1.2773 & 1.2745 \\

HBr & 1.4090 & 1.4095 & 1.4089 & 1.4095 & 1.4117 & 1.4123 & 1.4116 & 1.4122 & 1.4144 \\    

HI  & 1.6030 & 1.6033 & 1.6029 & 1.6031 & 1.6061 & 1.6067 & 1.6060 & 1.6066 & 1.6092\\
    
\end{tabular}
\end{ruledtabular}
\label{table:BL}
\end{table*}

\begin{table*}[ht!]
\caption{Comparison of harmonic vibrational frequency ($cm^{-1}$) calculated using the UCC-based method (qUCCSD and qUCCSD[T]) with the standard coupled cluster method (CCSD and CCSD(T)) under a fixed FNS and CD threshold $10^ {-5}$ with frozen core approximation.}
\begin{ruledtabular}
\begin{tabular}{cccccccccc}
Molecule & \multicolumn{2}{c}{CCSD} & \multicolumn{2}{c}{qUCCSD} & \multicolumn{2}{c}{CCSD(T)} &\multicolumn{2}{c}{qUCCSD[T]} & Exp.\cite{huber2013molecular_a} \\
\cline{2-3}
\cline{4-5}
\cline{6-7}
\cline{8-9}
& tz & qz & tz & qz& tz& qz & tz & qz \\ \hline

HF  &  4162.90  & 4182.53  & 4166.96 & 4187.37 & 4118.63 & 4135.74 &  4121.31 & 4138.25 & 4138.32  \\

HCl &  3011.63 &  3008.95  & 3011.42 & 3010.09 & 2987.34 & 2985.49 &  2991.43 & 2986.31 & 2990.94  \\
 
HBr &  2672.13  & 2696.76  & 2684.02 & 2694.06 & 2647.82 & 2669.06 &  2660.55 & 2678.72 & 2648.97 \\    

HI  &  2331.38  & 2341.71  & 2332.83 & 2341.81 & 2309.46 & 2298.01 &  2314.48 & 2320.16 & 2309.01\\
    
\end{tabular}
\end{ruledtabular}
\label{table:hvib}
\end{table*}

\subsection{Ionziation Potential and Electron Affinity }
To assess the performance of the qUCCSD[T] method for electronic properties sensitive to both correlation and relativistic effects, we calculated vertical ionization potentials (IP) and electron affinities (EA) for a series of hydrogen halides (HF, HCl, HBr, HI, HAt) and group 13 heavy elements (In, Tl, Nh), respectively. All calculations were performed using uncontracted aug-cc-pVTZ and aug-cc-pVQZ basis sets for lighter elements and dyall.ae3z/ae4z basis sets for heavier elements. The s-aug-dyall.ae3z/ s-aug-dyall.ae4z basis set has been used for the heavy elements in the EA calculations. The TIGHTFNS setting has been used for all the calculations. 
Table \ref{table:IP} shows that the qUCCSD[T] method yields IPs that are highly consistent with experimental data,\cite{banna1975photoelectron,cormack1997high,yencha1998threshold,rothe2013astatine,adam1992high} and often identical with CCSD(T) values. For example, the calculated IP of HCl using the qUCCSD[T] method is 12.718 eV (qz basis), very close to the experimental value of 12.74 ± 0.009 eV. Similar accuracy is observed for the heavier halides such as HBr and HI, confirming the efficacy of the relativistic framework. In almost all cases, the inclusion of partial triples correction improved upon the qUCCSD method, except for the HF molecule, where the results are of similar quality.

Table \ref{table:EA} presents EAs, where qUCCSD[T] again demonstrates good agreement with experimental references. For example, the EA of Tl calculated using qUCCSD[T] is 0.2828 eV (qz), compared to the experimental value of 0.32005 eV. While a slight underestimation is observed, the overall trend across In, Tl, and Nh is well-reproduced, highlighting the robustness of the qUCCSD[T] approach for systems with substantial spin–orbit coupling. The inclusion of the triples correction significantly improves over the qUCCSD results and gives a value comparable to the CCSD(T) method.These results underline the capacity of the qUCCSD[T] method to deliver balanced accuracy even for energy differences, extending its applicability beyond thermochemistry and structural predictions.

\begin{table*}[htbp]
\caption{Comparison of ionization potential (eV ) calculated using the UCC-based method (qUCCSD and qUCCSD[T]) with standard CC-based method (CCSD and CCSD(T)) under a fixed FNS and CD threshold $10^ {-5}$ with frozen core approximation. }
\begin{ruledtabular}
\begin{tabular}{cccccccccc}
Molecule & \multicolumn{2}{c} {CCSD} & \multicolumn{2}{c}{qUCCSD} & \multicolumn{2}{c}{CCSD(T)} & \multicolumn{2}{c}{qUCCSD[T]} & Exp. \\
\cline{2-3}
\cline{4-5}
\cline{6-7}
\cline{8-9}
    &   tz   &   qz   &   tz   &   qz   &   tz   &  qz    &   tz   &   qz \\ \hline
HF  & 16.043 & 16.106 & 16.038 & 16.101 & 16.137 & 16.200 & 16.135 & 16.198 & 16.12±0.04\cite{banna1975photoelectron}  \\
HCl & 12.551 & 12.639 & 12.552 & 12.640 & 12.625 & 12.719 & 12.624 & 12.718 & 12.74±0.009\cite{yencha1998threshold}  \\ 
HBr & 11.473 & 11.581 & 11.473 & 11.580 & 11.542 & 11.643 & 11.541 & 11.642 & 11.68±0.03\cite{adam1992high} \\    
HI  & 10.162 & 10.260 & 10.163 & 10.260 & 10.229 & 10.337 & 10.228 & 10.337 & 10.39±0.001\cite{cormack1997high} \\
HAt &  9.054 &  9.175 &  9.053 &  9.173 &  9.121 &  9.240 &  9.121 &  9.239 & 9.317\cite{rothe2013astatine}\\
    
\end{tabular}
\end{ruledtabular}
\label{table:IP}
\end{table*}

\begin{table*}[htbp]
\caption{Comparison of electron affinity (eV ) obtained using the UCC-based methods (qUCCSD and qUCCSD[T]) and the standard CC-based methods ($CCSD$ and $CCSD(T)$), with a fixed FNS and CD threshold of $10^ {-5}$ and $10^{-4.5}$ respectively. s-aug-dyall.ae3z (tz) and s-aug-dyall.ae4z (qz) basis sets were used for the calculations.}
\begin{ruledtabular}
\begin{tabular}{cccccccccc}
Atom & \multicolumn{2}{c}{CCSD} & \multicolumn{2}{c}{qUCCSD} & \multicolumn{2}{c}{CCSD(T)} & \multicolumn{2}{c}{qUCCSD[T]} & Exp. \\
\cline{2-3}
\cline{4-5}
\cline{6-7}
\cline{8-9}
   &   tz   &    qz  &   tz   &   qz   &  tz    &   qz   &   tz   & qz \\ \hline
In & 0.2074 & 0.2130 & 0.2082 & 0.2128 & 0.3152 & 0.3218 & 0.3140 & 0.3207 & 0.38392 \cite{walter2010electron}  \\
Tl & 0.1815 & 0.1958 & 0.1811 & 0.1950 & 0.2686 & 0.2836 & 0.2677 & 0.2828 & 0.32005\cite{walter2020electron}  \\
Nh & 0.5351 & 0.6114 & 0.5367 & 0.6137 & 0.6040 & 0.6849 & 0.6043 & 0.6849 & 0.776 \footnote{Theoretical best estimate (DC(B) +CCSDTQ+QED)\cite{guo2022relativistic}} \\
       
\end{tabular}
\end{ruledtabular}
\label{table:EA}
\end{table*}

\section{Conclusions}
We have developed and implemented a perturbative triples correction to the relativistic quadratic unitary coupled cluster singles and doubles method (qUCCSD), denoted as qUCCSD[T] .  The he X2CAMF Hamiltonian framework has been used to include relativistic effect in the calculations. The proposed method retains the Hermitian structure of the unitary ansatz while incorporating dynamic correlation effects from triple excitations in a computationally tractable manner. This formulation allows for efficient and accurate treatment of heavy-element systems where both relativistic and correlation effects play a significant role. The triples correction can be straightforwardly derived from the perturbation truncation of the qUCCSDT energy functional
benchmarking against Full CI and other UCC-based perturbative triples schemes in the non-relativistic regime demonstrates that qUCCSD[T] consistently yields the lowest deviation from exact results among existing triples correction schemes to unitary coupled cluster. In the relativistic domain, the method performs exceptionally well across a broad set of properties, including ligand dissociation enthalpies, equilibrium bond lengths, harmonic vibrational frequencies, ionization potentials, and electron affinities. In all cases, qUCCSD[T] matches the accuracy of the standard CCSD(T) method, offering a viable alternative with added advantage due to its  Hermitian nature.\\
The new triples correction scheme makes qUCC method as a promising and general-purpose approach for accurate quantum chemical calculations even on classical computers, especially in systems dominated by relativistic effects. Future work will focus on further scaling improvements, extension to excited states, ionized and electron-attached states using a propagator-like approach, and analytic calculation of properties. Work is in progress towards that direction.

\begin{acknowledgments}
The authors acknowledge support by the EU NextGenerationEU through the Recovery and Resilience Plan for Slovakia under project No. 09I03-03-V04-00117.
\end{acknowledgments}
\section{Supplementary Material}
The Supplementary Material contains the bond dissociation enthalpies for the 18 coinage metal-ligand complexes, as well as energies of the corresponding diatomic and atomic systems used to compute IP and EAs., calculated with FNS-CD-X2CAMF-based CCSD, qUCCSD, CCSD(T), qUCCSD[T] methods.

\section*{References}
\bibliographystyle{aipnum4-1}
\bibliography{aipsamp}
\end{document}